\documentclass[prl,preprint,amsmath,amssymb]{revtex4-1}

\usepackage{graphicx}
\usepackage{dcolumn}
\usepackage{bm}
\usepackage{multirow}
\usepackage{float}
\usepackage{wrapfig}
\setcitestyle{super}
\usepackage[normalem]{ulem}
\usepackage{color}
\usepackage{xr}
\usepackage{booktabs}
\usepackage{longtable}
\makeatletter
\providecommand \@ifxundefined [1]{%
 \@ifx{#1\undefined}
}%
\providecommand \@ifnum [1]{%
 \ifnum #1\expandafter \@firstoftwo
 \else \expandafter \@secondoftwo
 \fi
}%
\providecommand \@ifx [1]{%
 \ifx #1\expandafter \@firstoftwo
 \else \expandafter \@secondoftwo
 \fi
}%
\providecommand \href@noop [0]{\@secondoftwo}%
\providecommand \href [0]{\begingroup \@sanitize@url \@href}%
\providecommand \@href[1]{\@@startlink{#1}\@@href}%
\providecommand \@@href[1]{\endgroup#1\@@endlink}%
\providecommand \@sanitize@url [0]{\catcode `\\12\catcode `\$12\catcode
  `\&12\catcode `\#12\catcode `\^12\catcode `\_12\catcode `\%12\relax}%
\providecommand \@@startlink[1]{}%
\providecommand \@@endlink[0]{}%
\providecommand \url  [0]{\begingroup\@sanitize@url \@url }%
\providecommand \@url [1]{\endgroup\@href {#1}{\urlprefix }}%
\providecommand \urlprefix  [0]{URL }%
\providecommand \selectlanguage [0]{\@gobble}%
\providecommand \bibinfo  [0]{\@secondoftwo}%
\providecommand \bibfield  [0]{\@secondoftwo}%
\providecommand \BibitemShut  [1]{\csname bibitem#1\endcsname}%
\let\auto@bib@innerbib\@empty

\usepackage{ulem}
\usepackage{verbatim}
\usepackage{color}

\newcommand{\SPHIDE}[1]{{}}
\begin{document}
\preprint{}

\title{Role of Majorana Fermions in high-harmonic generation from Kitaev chain}

\author{Adhip Pattanayak}
\affiliation{%
Department of Physics, Indian Institute of Technology Bombay, Powai, Mumbai 400076, India }

\author{Sumiran Pujari}
\affiliation{%
Department of Physics, Indian Institute of Technology Bombay, Powai, Mumbai 400076, India }
                         
\author{Gopal Dixit}
\email[]{gdixit@phy.iitb.ac.in}
\affiliation{%
Department of Physics, Indian Institute of Technology Bombay, Powai, Mumbai 400076, India }




\begin{abstract}
The observation of Majorana fermions as collective excitations in condensed-matter 
systems
is an ongoing quest, and several state-of-the-art experiments have been performed
in the last decade. 
As a potential avenue in this direction,
we simulate the high-harmonic  spectrum 
of Kitaev's superconducting chain model
that hosts Majorana edge modes in its topological phase. 
It is well-known that this system exhibits a 
topological--trivial superconducting phase transition. 
We demonstrate that high-harmonic spectroscopy is sensitive to the phase transition
in presence of open boundary conditions due to the presence or absence of these edge modes.  
The population dynamics of the Majorana edge modes 
are different from the bulk modes,
which is the underlying reason for the distinct harmonic profile of both the phases. 
On the contrary, in presence of periodic boundary conditions with only bulk modes, 
high-harmonic spectroscopy becomes insensitive to the phase transition with
similar harmonic profiles in both phases. 
\end{abstract}

\maketitle 

Emission of radiation at prominently 
higher integer multiple frequencies  
of the incident laser frequency due to strong nonlinear interaction of intense laser fields 
with matter is known as high-harmonic generation (HHG)~\cite{ghimire2019, kruchinin2018colloquium, ghimire2011observation}. 
The emitted radiation  encodes the information about laser-driven 
sub-cycle electron dynamics which forms the basis of 
HHG spectroscopy.  
In recent years, HHG in solids has become a method of choice to probe various
aspects of condensed matter such as the observation of Bloch oscillations~\cite{schubert2014sub, mcdonald2015interband}, 
examining  the dynamics of the defects in solids~\cite{mrudul2020high, pattanayak2020influence},  band structure 
tomography~\cite{vampa2015all, lanin2017mapping, tancogne2017impact}, observation of the valley pseudospin~\cite{langer2018lightwave, mrudul2021light, mrudul2021controlling}, 
imaging valence electrons~\cite{lakhotia2020laser, pattanayak2019direct},  
realisation of petahertz currents in solids~\cite{luu2015extreme, garg2016multi}, probing Berry phases~\cite{banks2017dynamical, luu2018measurement}, 
and phase transitions driven by both light and topology~\cite{bauer2018high, jurss2019high, murakami2018high, reimann2018subcycle, imai2020high, baykusheva2020strong, bai2020high, borsch2020super, mrudul2020high_new}.  

In the field of strongly correlated matter, 
topological superconductivity and associated phase transitions is presently 
one of the major topics~\cite{beenakker2013search, sato2017topological} 
as it is associated with the emergence 
of the Majorana fermion -- particles that are their own antiparticles~\cite{majorana}. 
The notion of  the Majorana fermion is also of great importance
in nuclear and particle physics, apart from solid-state physics~\cite{elliott2015colloquium}. 
In a condensed matter context, Majorana fermions  
are emergent quasiparticle excitations that
have been proposed as the physical basis to realize
qubits that are robust against decoherence 
and fault-tolerant topological quantum computation
~\cite{kitaev2003fault, alicea2012new, nayak2008non, sarma2015majorana}.  
In this work, we discuss  
the role of the  Majorana fermions  
by probing the topological--trivial superconducting phase transition using HHG spectroscopy.

Experimental observation of Majorana fermions is an ongoing quest, and 
several experiments on semiconductor nanowires, magnetic chains and superconductors, 
using transport and electrical measurements with scanning tunneling spectroscopy and Coulomb-blockade spectroscopy, have been 
reported~\cite{mourik2012signatures, das2012zero, albrecht2016exponential, kim2018toward, wang2018evidence, jack2019observation, lutchyn2018majorana, wang2020evidence}. 
Topological superconductors are considered as a promising 
platform for Majorana fermions since the original proposal of Kitaev.~\cite{kitaev2001unpaired}
Thus for this study, 
we choose the paradigmatic 
one dimensional (1D) Kitaev chain of spinless fermions as
a suitable  model system. 
The many-body Hamiltonian for this 1D superconducting chain is
\begin{eqnarray}\label{eq:kitaev_model}
\mathcal{H}  & = &  -\mu \sum_{j=1}^{L}c_{j}^{\dagger}c_{j}-t_0\sum_{j=1}^{L}\left ( c_{j+1}^{\dagger}c_{j} + c_{j}^{\dagger}c_{j+1} \right ) \nonumber \\
&&  +\Delta\sum_{j=1}^{L}\left ( c_{j}c_{j+1}+c_{j+1}^{\dagger}c_{j}^{\dagger} \right ).
\end{eqnarray} 
Here, $\mu$ is the onsite energy, $t_0$ is a nearest neighbour hopping term, and 
$\Delta$  is an order parameter (chosen to be purely real). 
The term with $\Delta$ is  a superconducting pairing term in mean-field approximation,
 which creates or annihilates pairs of particles at neighbouring lattice sites. 
$c_j^\dagger$ ($c_j$) is fermionic creation (annihilation) operator at site $j$ and $L$ represents the number of lattice sites. 

In the appropriate region of parameter space,
the system described by Eq.~(\ref{eq:kitaev_model}) 
can host a topological superconductor that leads to 
the emergence of Majorana zero-energy edge modes 
(MZMs)~\cite{elliott2015colloquium, jack2019observation, lutchyn2018majorana}.  
The boundary of the phase transition from the topological to the 
trivial superconducting side depends on 
the  critical values of $\mu$, $t_0$ and $\Delta$~\cite{beenakker2013search, elliott2015colloquium}.
For example, for $\Delta = t_0$, 
the system is in the trivial phase when $\left | \mu/t_0\right | > 2$   
in the limit $L\rightarrow \infty$. 
This is qualitatively similar to the limit $\left | \mu/t_0\right |\rightarrow \infty$. 
For $\left | \mu/t_0\right | < 2$, the system behaves as  a topological superconductor
with boundary MZMs in presence of edges. 
At $\left | \mu/t_0\right | = 2$, the gap closes 
which causes the boundary zero modes to hybridize and disappear~\cite{alicea2012new}.

\begin{figure}[h!]
\begin{center}
\includegraphics[width=\linewidth]{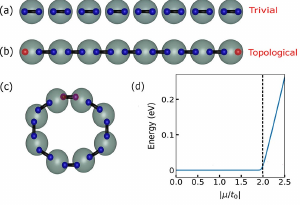}
\end{center}
\caption{ {\bf Notion of Majorana fermions in one dimensional chain.} Two phases of the  
1D superconducting Kitaev  chain with open boundary: 
(a) the trivial phase has Majorana fermions (blue spheres)
bound in pairs located on the same site of the physical lattice, represented by grey spheres. 
The other phase is (b) the topological phase, 
where Majorana fermions are bound in pairs located 
on the neighbouring sites leading to two unpaired Majorana fermions 
at both ends of the chain, represented by the red spheres. 
(c) The two unpaired Majorana at  the ends of the chain again become bound in a pair when 
periodic or closed boundary  condition is employed. 
(d) Lowest many-body excitation energy for $L = 128$ as a function of  $|\mu/t_{0}|$.}  
\label{fig1}
\end{figure}

To set the stage for the HHG study, we briefly 
recall this topological--trivial phase structure 
and its association with the MZMs.
Following Ref.~\cite{kitaev2001unpaired}, 
one rewrites Eq.~(\ref{eq:kitaev_model}) in terms of Majorana operators
$c^\dagger_j = \frac{1}{2}(\gamma_{j,1} + i \gamma_{j,2})$ and
$c_j = \frac{1}{2}(\gamma_{j,1} - i \gamma_{j,2})$
to arrive at 
$\mathcal{H} = \frac{i}{2} \Big[ -\mu \sum_{j=1}^{L} \gamma_{j,1} \gamma_{j,2}
+ (-t_0 + \Delta) \sum_{j=1}^{L} \gamma_{j,1} \gamma_{j+1,2}   
+ (t_0 + \Delta) \sum_{j=1}^{L} \gamma_{j,2} \gamma_{j+1,1} \Big]$
up to a constant.
The system is in the trivial phase when two Majorana fermions are bound on the same site  
as sketched in Fig~\ref{fig1}(a). This is clearly seen when $\mu$ is the dominant
scale in $\mathcal{H}$ as expressed in terms of the Majorana operators.
On the other hand, two unpaired Majorana fermions emerge 
at the boundary when 
Majorana fermions are bound in pairs 
on the neighbouring sites as sketched in Fig~\ref{fig1}(b) 
that corresponds with the system being in the topological superconducting phase. 
This is easiest to see for $t_0 = \Delta$ and $\mu=0$.
More generally in the topological phase,
it is understood that the two unpaired Majorana fermions (or MZMs) are 
exponentially localised at each physical boundary~\cite{kitaev2001unpaired}. 
This notion of emergent unpaired Majorana fermions at the boundary
or edge, expectedly, gets lost  when 
closed (periodic)  boundary condition (PBC) is employed as sketched in Fig~\ref{fig1}(c). 
The topological--trivial transition can be seen by tracking 
the lowest many-body excitation energy as a function of  
$\left | \mu/t_0\right |$ for the open chain as shown in Fig~\ref{fig1}(d). 
On the topological side, there is an effectively zero energy (fermionic) excitation
which is composed from the unpaired Majorana modes localised at the opposite edges
as in Fig.~\ref{fig1}(b), whereas on the trivial side, such zero energy modes are lost.

\begin{figure}[h]
\begin{center}
\includegraphics[width=\linewidth]{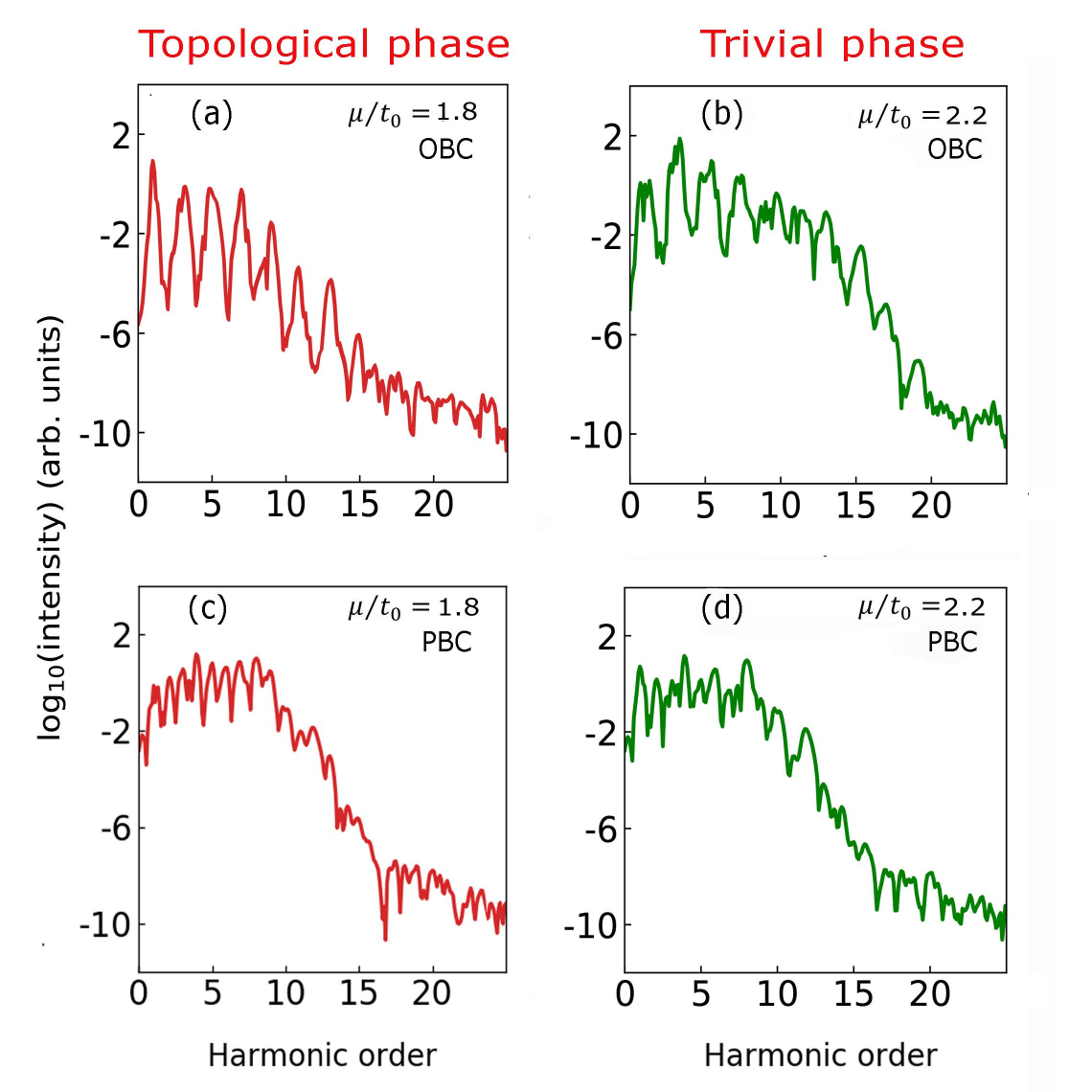}
\end{center}
\caption{ {\bf High-harmonic spectrum as a function of $\mu/t_{0}$ with different boundary conditions.} High-harmonic spectra on the two side of the phase transition 
of the 1D superconducting Kitaev chain. 
The spectra for (a) the topological phase at $\mu/t_{0} = 1.8$, and 
(b) the trivial  phase at $\mu/t_{0} = 2.2$ when the open boundary condition (OBC) is used. 
The spectra  (c) and (d) are same as (a) and (b), respectively, with 
periodic boundary condition (PBC) imposed on the Kitaev chain. 
$L = 128$ is used while simulating the harmonic spectra. 
A linearly polarised laser pulse with a peak amplitude of 30 {MV/m} and  9.1 $\mu$m wavelength having ten optical cycles with sine squared envelope is used to obtain the spectrum.} 
\label{fig2}
\end{figure} 

High-harmonic spectra for the 1D Kitaev chain with open boundary conditions 
(OBC) and PBC are presented in Fig.~\ref{fig2}. 
The left and right panels in the figure correspond to topological and trivial phases, respectively,
on either side of the transition at $ \mu/t_0 = 2.0$.  
To observe the sensitivity of  HHG spectroscopy on the two phases, we have chosen 
the values of  $ \mu/t_0$ close to transition point. 
A gradual decrease in the harmonic intensity up to 15$^{\textrm{th}}$ harmonics  is visible 
when the system  is in the topological phase  at $ \mu/t_0 = 1.8$ [see Fig.~\ref{fig2}(a)]. 

\begin{figure}[]
\begin{center}
\includegraphics[width=\linewidth]{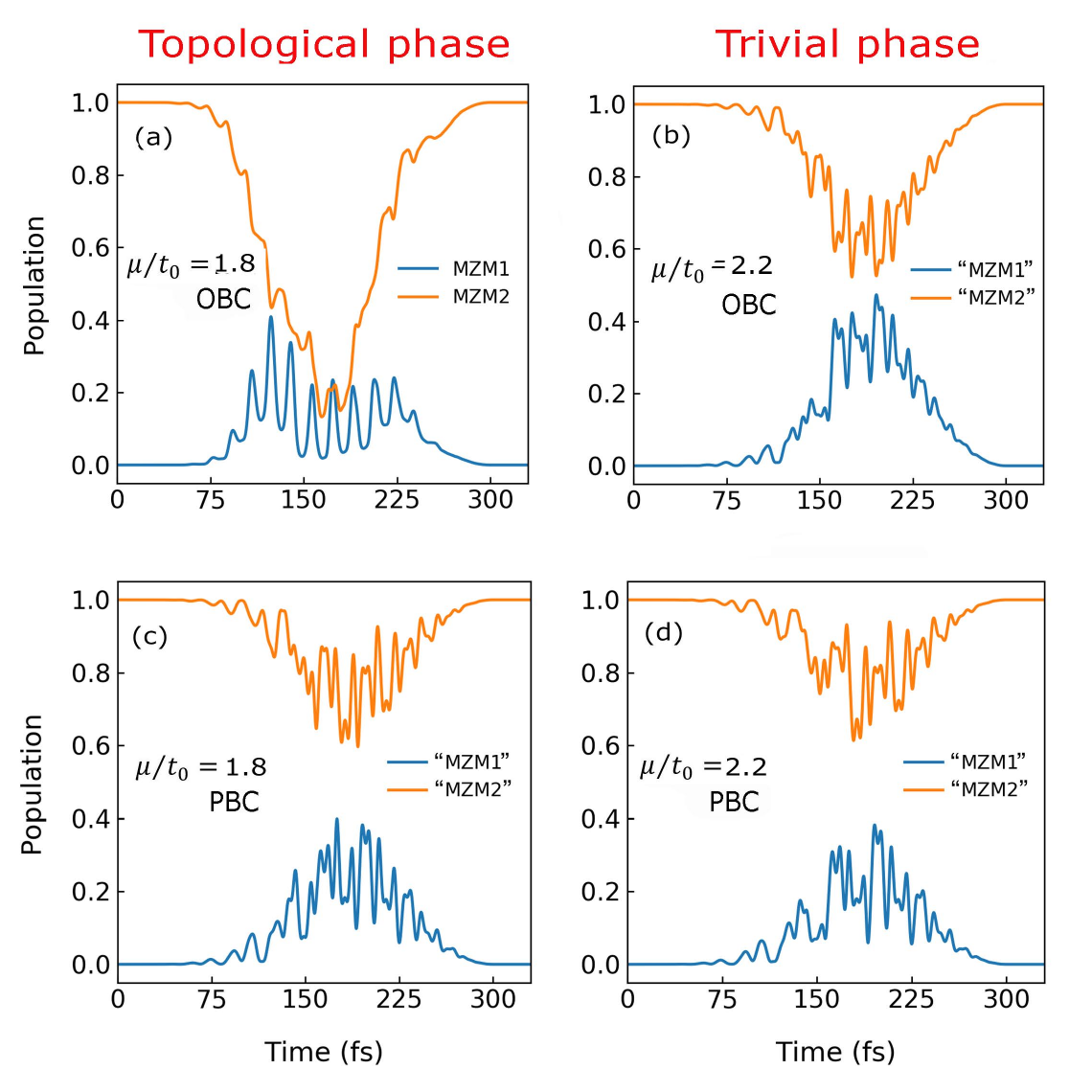}
\end{center}
\caption{ {\bf Population dynamics of Majorana zero modes in the presence of strong laser pulse.} Population dynamics during the laser 
pulse for  (a) doubly degenerate Majorana zero-energy modes (MZM1 and MZM2 in the legend) 
in the BdG spectrum on the topological side at $ \mu/t_0 = 1.8$ and 
(b)  the corresponding 
non-zero non-degenerate modes (which we have still called ``MZM1" and ``MZM2"
to avoid extra notation)
on the trivial side at $ \mu/t_0 = 2.2$ with OBC.
The population dynamics (c) and (d) are same as (a) and (b), respectively with PBC
where none of the modes are at zero energy due to the absence of MZMs.} 
\label{fig3}
\end{figure} 

However, the spectrum for the trivial phase at $ \mu/t_0 = 2.2$ is significantly  
different from the topological phase, as 
evident from the right panel of the figure [see Fig.~\ref{fig2}(b)].   
In this case, the extent of the harmonics  is relatively broad, 
ranging from 3$^{\textrm{rd}}$ to  15$^{\textrm{th}}$ order. 
The intensity of  the harmonics in the plateau region is  comparable,
and the spectra extend up to the 17$^{\textrm{th}}$ harmonic. 
In both the phases, the harmonic spectra exhibit  odd harmonics only
as expected from the inversion symmetry present in the 1D system.
Note that energy  eigen-spectra are almost same at   $ \mu/t_0 = 1.8$ and 2.2 as evident from
Fig.~S1.  
The band gap at $ \mu/t_0 = 1.8$ is $\sim$0.109 eV (if MZM states at zero energy are excluded), 
and at $ \mu/t_0 = 2.2$ is  $\sim$0.11 eV.  
In  Fig.~S7, we also present a series of harmonic spectra for different values of  $ \mu/t_0$ 
(= 1.6, 1.7, 1.8, 1.9, 2.1, 2.2, 2.3, 2.4)
across the transition  through which their differences on the two sides of the transition becomes evident.

 \begin{figure}[h]
\begin{center}
\includegraphics[width=\linewidth]{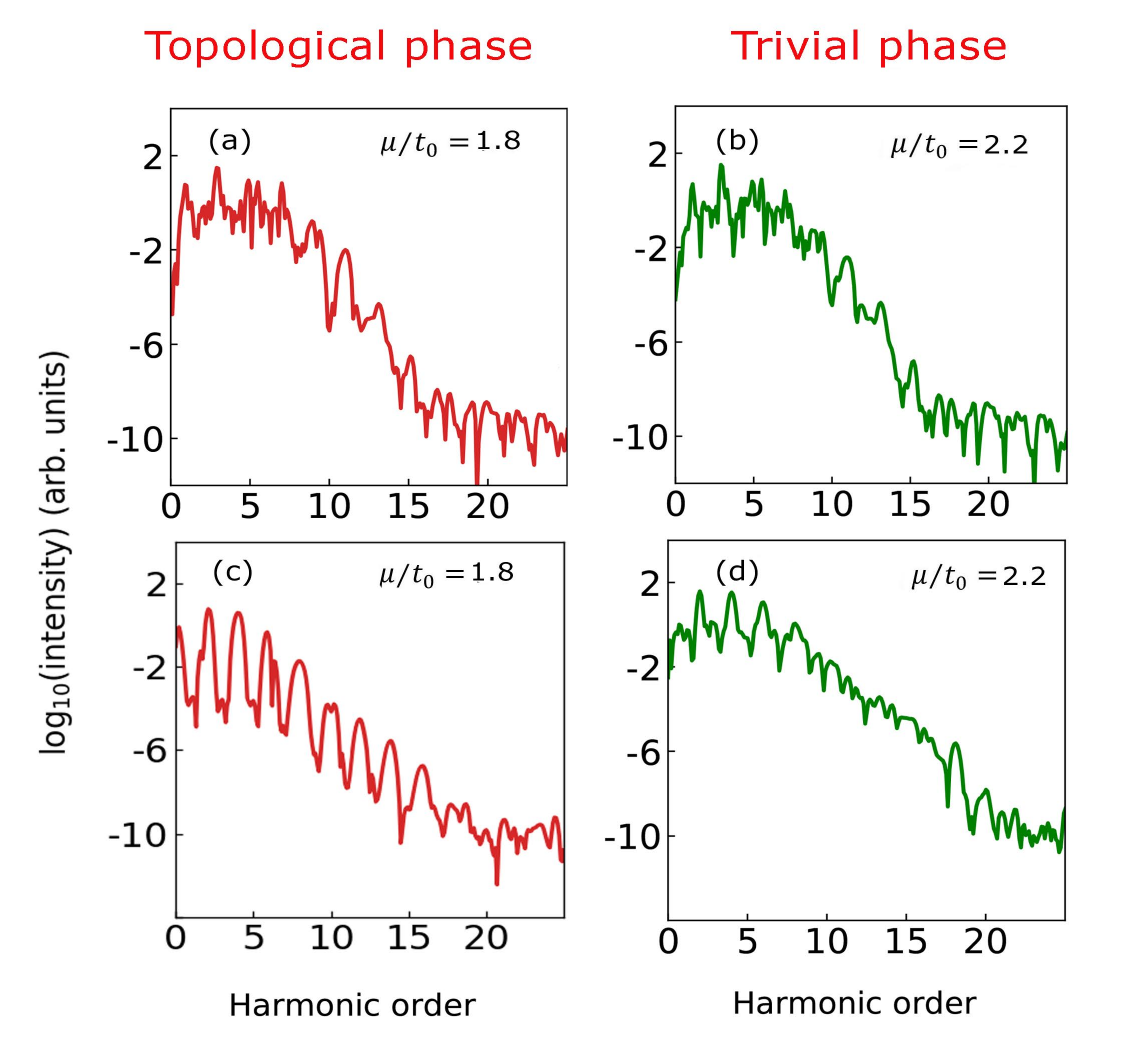}
\end{center}
\caption{ {\bf High-harmonic spectrum with and without Majorana zero modes of the 1D superconducting Kitaev chain with  open boundary condition (OBC).}
Harmonic spectrum where for both occupied and unoccupied MZM states are not considered for 
(a) topological phase $\mu/t_{0} = 1.8$, and  (b) trivial superconducting phase $\mu/t_{0} = 2.2$.  Harmonic spectrum originating only from occupied MZM state  for 
(c) topological phase $\mu/t_{0} = 1.8$, and  (d) trivial superconducting phase $\mu/t_{0} = 2.2$. 
A linearly polarised laser pulse with a peak amplitude of 30 {MV/m} and  9.1 $\mu$m wavelength having ten optical cycles with sine squared envelope is used to obtain the spectrum.} 
\label{fig4}
\end{figure} 

The situation becomes very different when PBC is imposed on 
the 1D Kitaev chain such that there are no boundaries. 
In this case, the harmonic spectra corresponding to $ \mu/t_0 = 1.8$ and 2.2 are very similar as 
reflected from Figs.~\ref{fig2}(c) and \ref{fig2}(d). 
The small differences between both the spectra 
arise due to slight changes in the eigen-spectra  at $ \mu/t_0 = 1.8$ and 2.2. 
The analysis of Fig.~\ref{fig2} concludes that 
the harmonic spectra  are very similar when PBC is employed, 
whereas markedly different for both the phases when OBC is employed. 
This indicates that the MZMs are playing a distinct 
role in the harmonic generation as seen by the contrast in the
HHG spectra in both the phases with different boundary conditions. 
In the following, we will investigate 
how the MZMs are associated with the different harmonic spectra  in 
the two  phases. 

To elucidate the role of the MZMs in the harmonic generation, population dynamics 
of the doubly degenerate MZMs in Bogoliubov-de Gennes (BdG) spectrum on the topological side 
 is presented in Fig.~\ref{fig3}. 
We recall that  the degeneracy of the  two zero modes is lifted in the trivial phase,
whereby they lose their Majorana character. 
When compared to these corresponding non-degenerate and non-zero modes in 
the trivial phase, 
the population dynamics during the laser pulse
on the topological side at $ \mu/t_0 = 1.8$ shows that the 
degenerate Majorana states show drastically different behaviours 
[see Figs.~\ref{fig3}(a) and ~\ref{fig3}(b)].   
This noticeable difference in population dynamics in both the phases 
owing to the topologically enforced
zero mode character of the Majorana excitations or lack thereof can be 
correlated with the different  harmonic spectra  observed
in both the phases as shown in Fig.~\ref{fig2}.

In presence of PBC, when there are no MZMs present, 
the temporal evolution of the corresponding modes are quite similar 
at $ \mu/t_0 = 1.8$ and 2.2  
[see Figs.~\ref{fig3}(c) and ~\ref{fig3}(d)]. 
This similarity in their population evolution 
indicates that both the modes 
are contributing in the same way to  the harmonic generation at $ \mu/t_0 = 1.8$ and 2.2;  
this correlates with almost the same harmonic spectra for both the phases with PBC 
as shown in Figs.~\ref{fig2}(c) and ~\ref{fig2}(d) 
in contrast to the above discussion on the previous case with OBC. 
Moreover, their behaviour is similar
to the OBC case on the trivial side [Fig.~\ref{fig3}(b)] when these modes have become part
of the bulk and not localized at the boundary anymore.
We recall again that the notion of the unpaired Majorana fermions exists only 
on the topological side with OBC which is in line with the preceding discussions.
The present findings thus establish that the 
harmonic spectra and the population dynamics of the MZMs are different for both the phases 
for the 1D Kitaev chain with OBC.  This
concludes that high-harmonic spectroscopy is sensitive to the presence of  the MZMs. 
The notion of unpaired Majorana fermion does not exist 
for the 1D Kitaev chain with PBC.
As a result,  the harmonic spectra and  the population dynamics 
are similar for different values of $ \mu/t_0$. 

In order to highlight further 
the role of MZMs in the differences of the harmonic profiles in presence of OBC, 
we also show the harmonic spectra obtained by artificially suppressing the presence of MZMs in the eigen-spectra. 
The harmonic spectra at  $ \mu/t_0 = 1.8$ and 2.2 with OBC 
are almost same as evident from Figs.~\ref{fig4}(a) and  ~\ref{fig4}(b). 
The minor difference in the spectra are originating from the minor difference in the eigen-spectra. 
Moreover, the harmonic spectra solely originating from one occupied MZM at  $ \mu/t_0 = 1.8$ and
corresponding non-zero mode at  $ \mu/t_0 = 2.2$ is shown in Figs.~\ref{fig4}(c) and  ~\ref{fig4}(d). 
As reflected from the figure, the spectra are drastically different from each other. 
Furthermore, the population dynamics of the highest occupied and lowest unoccupied eigen-states (representative of the bulk
modes)
corresponding to the spectra in Figs.~\ref{fig4}(a) and  ~\ref{fig4}(b) are similar as expected (see Fig.~S8). 
On the other hand, the population dynamics corresponding to spectra in Figs.~\ref{fig4}(c) and  ~\ref{fig4}(d) are drastically different (see Fig.~S8).
Results presented in Fig.~\ref{fig4}  unequivocally establish that the similarities and differences 
in the spectra under different situations are due to the presence or absence of 
the MZM modes.  
Also,  the system  continues to exhibit different harmonic profiles
in the two phases (for OBC)
for different laser intensities (see Figs.~S3 and S4) 
and wavelengths  (see Figs.~S5 and S6).

We now briefly comment on the domain of validity for the observations made earlier.
Keeping our choice of the laser parameters fixed, we find that
a variation of $\sim 11\%$ in the $t_0=\Delta$ scale does not change our
conclusions. Keeping the $t_0$ scale fixed, a variation of $~\sim \pm 8\%$ in 
$\Delta/t_0$ keeps conclusions unchanged as well. Beyond these variations,
the laser parameters have to be tuned accordingly to find the suitable values.
Thus, analogous simulations can be helpful in determining 
the appropriate laser characteristics when applying HHG to different physical
systems. Finally, it is expected that the HHG obtained from the time evolution of the
many-body Hamiltonian in Eq.~(\ref{eq:kitaev_model})  and 
from that of the BdG-recasted ``one-particle" 
Hamiltonian should be the same in principle. 
We have confirmed this expectation by explicit 
numerical simulations (results not shown here), which is an
independent check on the accuracy of our numerical time evolution procedures.

In summary, our work demonstrates that HHG spectroscopy is 
sensitive to  the topological--trivial phase transition in 1D superconducting Kitaev chain. 
The extent of the harmonic plateau and energy cutoff 
in the trivial phase are distinguishable from the harmonic spectra in the topological phase. 
This distinguishability is further consolidated by various harmonic spectra for different intensities and wavelengths of driving laser as presented in supplementary material.
The population dynamics of the two Majorana states 
and their contributions to the total harmonic spectra
are very different for both the phases, which we established
to be correlated
with the differences in the HHG spectra of the two phases. 
This distinct signature may potentially point to a new probe of Majorana
modes in solids.
A physical explanation for the numerically observed differences in
the population dynamics of MZMs and bulk modes that were shown to underlie the
HHG spectra is desirable for better understanding.
Finally, the feasibility of coupling existing realisations of topological
superconductivity with potential MZMs
 to \emph{intense} laser fields in the laboratory 
is an open issue~\cite{Ohm_Hassler_2014}, and we hope 
our work stimulates attempts in this direction.

\section{Methods}
To probe the topological--trivial phase transition by HHG spectroscopy, 
we imagine that the 1D Kitaev chain is exposed to an ultrashort, strong laser pulse. 
For simulating this process, it is 
useful to recast the Hamiltonian, given in Eq.~(\ref{eq:kitaev_model}), 
in the two-component Nambu-Bogoliubov-de Gennes (BdG) formalism.~\cite{DeGennes_book} 
The Hamiltonian of Eq.~(\ref{eq:kitaev_model}) 
can be written as $\mathcal{H} = \frac{1}{2}\;C^{\dagger}\;H_{\textrm{BdG}}\;C$
with  $C = (c_1,...,...,c_L,c^{\dagger}_1,...,...,c^{\dagger}_L)^T$  as a column vector 
containing all fermionic creation and annihilation operations.  
The one-particle matrix elements contained in $H_{\textrm{BdG}}$ 
can be compactly written as 
\begin{eqnarray}\label{eq:bdg_kitaev_model}
H_{\textrm{BdG}} & = & -\mu \sigma_{z}\sum_{n}\left | n \right \rangle\left \langle n \right | \nonumber \\  
&&  - \sum_{n} [(t_{0} \sigma_z -  i \Delta \sigma_y) \left | n \right \rangle\left \langle n+1 \right | + \textrm{h.c.}]
\end{eqnarray}
by using Pauli matrices $\sigma$.
$H_{\textrm{BdG}}$ thus 
has the dimension of $2L \times 2L$, where 
$|n\rangle$ corresponds to the $n$-th 
site of the chain and $\textrm{h.c.}$ stands for the hermitian conjugate. Here,  
$\sigma$ operates on
the particle and hole indices of the BdG formalism. 

In presence of the laser pulse, 
the hopping term now acquires a time-dependent Peierls phase $\theta(t)$. 
The time derivative of $\theta(t)$ is directly proportional to the applied electric field as 
$E(t) = -\frac{1}{a_0}\frac{d\theta(t)}{dt} = -\frac{dA(t)}{dt}$ with 
$a_0$  as the lattice constant and $A(t)$ as the time-dependent vector potential of the driving laser pulse. 
Here, we are thus assuming a spatially constant field due to the
laser. This assumption corresponds to the limit when the wavelength of the laser
is much larger than the 1D system under probe 
which will govern our choice of laser wavelength
subsequently. 
Also, the value of $\Delta$ changes as a function of time in the presence of laser 
(see Fig.~S2). To simulate the time-dependent changes in (real) $\Delta$, 
we have used the fact $\Delta(t) \propto \langle c_j c_{j+1} \rangle(t)$ for the
pairing term as time progresses,
e.g. see Ref.~\cite{zachmann2013ultrafast}. 
Time-dependent Schrodinger equation for the modes of the
BdG-recasted Hamiltonian  
is numerically solved to simulate the high-harmonic spectrum along 
with time-evolving $\Delta(t)$ -- i.e., at each (small) discrete time step,
the BdG modes are first computed using the Hamiltonian parameters from the previous
time step, and then the pairing term coefficient 
$\Delta$ is re-computed using these new set of BdG modes just computed
to supply the Hamiltonian parameters for the next time step. One ensures the convergence
of this procedure with respect to the size of the discretized time steps which 
simulates the continuous time evolution.

The high-harmonic power 
spectrum is obtained by the modulus square of the Fourier-transform of the total current 
$j(t) = \sum_k\left \langle \psi_k(t)|\hat{J}(t)|\psi_k(t) \right \rangle$ where 
\begin{equation}
\hat{J}(t) = -ie a_{0} t_{0}\sigma_z \sum_{n} [\mathrm{e}^{-i\theta(t)} \left | n \right \rangle\left \langle n+1 \right | - \textrm{h.c.}],
 \end{equation} 
and $|\psi_k(t) \rangle$ is the time-propagated state at time $t$. 
Half of the states in the BdG spectrum make up the many-body ground
state which is considered as the initial state for the time evolution
in order to obtain the harmonic spectra. On the topological side, 
this corresponds to (lower) one of the two zero-energy modes in the
BdG spectrum being occupied.  

In this work,  $t_{0}$ = 0.26 eV and $a_{0} \sim$ 0.4 nm are 
considered for the simulations. In what follows, we set $\Delta = t_{0}$ 
at the initial time instant to
describe our main observations. 
A linearly polarised laser pulse of 9.1 $\mu$m wavelength 
having ten optical cycles with sine squared envelope is used. 
The polarisation of the driving pulse  is along the 1D Kitaev chain and 
has a peak amplitude of 30 {MV/m}.  
Choice of the laser parameters is motivated from Ref.~\cite{silva2018high}.   
Time-step of  0.1 atomic unit ($\sim$ 2.5 as) is used for time propagation.

\section*{Data Availability}
Data that support the plots within this paper and other findings of this study are available from the corresponding authors upon reasonable request.

\section*{Code Availability}
Code that support the findings of this study are available from the corresponding authors on reasonable request.

\section*{Acknowledgements}
A. P. acknowledges fruitful discussion with Souvik Bandyopadhyay from IIT Kanpur,   
and sandwich doctoral fellowship from Deutscher Akademischer Austauschdienst (DAAD, reference no. 57440919). 
SP acknowledges financial support from Science and Engineering Research Board (SERB) 
India (SRG/2019/001419).
G. D. acknowledges financial support from Science and Engineering Research Board (SERB) India 
(Project No. ECR/2017/001460). 

\section*{Competing Interests}
The authors declare no competing interests. 

\newpage

\begin{thebibliography}{10}
\expandafter\ifx\csname url\endcsname\relax
  \def\url#1{\texttt{#1}}\fi
\expandafter\ifx\csname urlprefix\endcsname\relax\def\urlprefix{URL }\fi
\providecommand{\bibinfo}[2]{#2}
\providecommand{\eprint}[2][]{\url{#2}}

\bibitem{ghimire2019}
\bibinfo{author}{Ghimire, S.} \& \bibinfo{author}{Reis, D.~A.}
\newblock \bibinfo{title}{High-harmonic generation from solids}.
\newblock \emph{\bibinfo{journal}{Nature Physics}}
  \textbf{\bibinfo{volume}{15}}, \bibinfo{pages}{10--16}
  (\bibinfo{year}{2019}).

\bibitem{kruchinin2018colloquium}
\bibinfo{author}{Kruchinin, S.~Y.}, \bibinfo{author}{Krausz, F.} \&
  \bibinfo{author}{Yakovlev, V.~S.}
\newblock \bibinfo{title}{Colloquium: Strong-field phenomena in periodic
  systems}.
\newblock \emph{\bibinfo{journal}{Reviews of Modern Physics}}
  \textbf{\bibinfo{volume}{90}}, \bibinfo{pages}{021002}
  (\bibinfo{year}{2018}).

\bibitem{ghimire2011observation}
\bibinfo{author}{Ghimire, S.} \emph{et~al.}
\newblock \bibinfo{title}{Observation of high-order harmonic generation in a
  bulk crystal}.
\newblock \emph{\bibinfo{journal}{Nature Physics}}
  \textbf{\bibinfo{volume}{7}}, \bibinfo{pages}{138--141}
  (\bibinfo{year}{2011}).

\bibitem{schubert2014sub}
\bibinfo{author}{Schubert, O.} \emph{et~al.}
\newblock \bibinfo{title}{Sub-cycle control of terahertz high-harmonic
  generation by dynamical bloch oscillations}.
\newblock \emph{\bibinfo{journal}{Nature Photonics}}
  \textbf{\bibinfo{volume}{8}}, \bibinfo{pages}{119} (\bibinfo{year}{2014}).

\bibitem{mcdonald2015interband}
\bibinfo{author}{McDonald, C.~R.}, \bibinfo{author}{Vampa, G.},
  \bibinfo{author}{Corkum, P.~B.} \& \bibinfo{author}{Brabec, T.}
\newblock \bibinfo{title}{Interband bloch oscillation mechanism for
  high-harmonic generation in semiconductor crystals}.
\newblock \emph{\bibinfo{journal}{Physical Review A}}
  \textbf{\bibinfo{volume}{92}}, \bibinfo{pages}{033845}
  (\bibinfo{year}{2015}).

\bibitem{mrudul2020high}
\bibinfo{author}{Mrudul, M.~S.}, \bibinfo{author}{Tancogne-Dejean, N.},
  \bibinfo{author}{Rubio, A.} \& \bibinfo{author}{Dixit, G.}
\newblock \bibinfo{title}{High-harmonic generation from spin-polarised defects
  in solids}.
\newblock \emph{\bibinfo{journal}{npj Computational Materials}}
  \textbf{\bibinfo{volume}{6}}, \bibinfo{pages}{1--9} (\bibinfo{year}{2020}).

\bibitem{pattanayak2020influence}
\bibinfo{author}{Pattanayak, A.}, \bibinfo{author}{Mrudul, M.~S.} \&
  \bibinfo{author}{Dixit, G.}
\newblock \bibinfo{title}{Influence of vacancy defects in solid high-order
  harmonic generation}.
\newblock \emph{\bibinfo{journal}{Physical Review A}}
  \textbf{\bibinfo{volume}{101}}, \bibinfo{pages}{013404}
  (\bibinfo{year}{2020}).

\bibitem{vampa2015all}
\bibinfo{author}{Vampa, G.} \emph{et~al.}
\newblock \bibinfo{title}{All-optical reconstruction of crystal band
  structure}.
\newblock \emph{\bibinfo{journal}{Physical Review Letters}}
  \textbf{\bibinfo{volume}{115}}, \bibinfo{pages}{193603}
  (\bibinfo{year}{2015}).

\bibitem{lanin2017mapping}
\bibinfo{author}{Lanin, A.~A.}, \bibinfo{author}{Stepanov, E.~A.},
  \bibinfo{author}{Fedotov, A.~B.} \& \bibinfo{author}{Zheltikov, A.~M.}
\newblock \bibinfo{title}{Mapping the electron band structure by intraband
  high-harmonic generation in solids}.
\newblock \emph{\bibinfo{journal}{Optica}} \textbf{\bibinfo{volume}{4}},
  \bibinfo{pages}{516--519} (\bibinfo{year}{2017}).

\bibitem{tancogne2017impact}
\bibinfo{author}{Tancogne-Dejean, N.}, \bibinfo{author}{M{\"u}cke, O.~D.},
  \bibinfo{author}{K{\"a}rtner, F.~X.} \& \bibinfo{author}{Rubio, A.}
\newblock \bibinfo{title}{Impact of the electronic band structure in
  high-harmonic generation spectra of solids}.
\newblock \emph{\bibinfo{journal}{Physical Review Letters}}
  \textbf{\bibinfo{volume}{118}}, \bibinfo{pages}{087403}
  (\bibinfo{year}{2017}).

\bibitem{langer2018lightwave}
\bibinfo{author}{Langer, F.} \emph{et~al.}
\newblock \bibinfo{title}{Lightwave valleytronics in a monolayer of tungsten
  diselenide}.
\newblock \emph{\bibinfo{journal}{Nature}} \textbf{\bibinfo{volume}{557}},
  \bibinfo{pages}{76} (\bibinfo{year}{2018}).

\bibitem{mrudul2021light}
\bibinfo{author}{Mrudul, M.~S.}, \bibinfo{author}{Jim{\'e}nez-Gal{\'a}n,
  {\'A}.}, \bibinfo{author}{Ivanov, M.} \& \bibinfo{author}{Dixit, G.}
\newblock \bibinfo{title}{Light-induced valleytronics in pristine graphene}.
\newblock \emph{\bibinfo{journal}{Optica}} \textbf{\bibinfo{volume}{8}},
  \bibinfo{pages}{422--427} (\bibinfo{year}{2021}).

\bibitem{mrudul2021controlling}
\bibinfo{author}{Mrudul, M.} \& \bibinfo{author}{Dixit, G.}
\newblock \bibinfo{title}{Controlling valley-polarisation in graphene via
  tailored light pulses}.
\newblock \emph{\bibinfo{journal}{Journal of Physics B}}
  \textbf{\bibinfo{volume}{54}}, \bibinfo{pages}{224001}
  (\bibinfo{year}{2021}).

\bibitem{lakhotia2020laser}
\bibinfo{author}{Lakhotia, H.} \emph{et~al.}
\newblock \bibinfo{title}{Laser picoscopy of valence electrons in solids}.
\newblock \emph{\bibinfo{journal}{Nature}} \textbf{\bibinfo{volume}{583}},
  \bibinfo{pages}{55--59} (\bibinfo{year}{2020}).

\bibitem{pattanayak2019direct}
\bibinfo{author}{Mrudul, M.~S.}, \bibinfo{author}{Pattanayak, A.},
  \bibinfo{author}{Ivanov, M.} \& \bibinfo{author}{Dixit, G.}
\newblock \bibinfo{title}{Direct numerical observation of real-space
  recollision in high-order harmonic generation from solids}.
\newblock \emph{\bibinfo{journal}{Physical Review A}}
  \textbf{\bibinfo{volume}{100}}, \bibinfo{pages}{043420}
  (\bibinfo{year}{2019}).

\bibitem{luu2015extreme}
\bibinfo{author}{Luu, T.~T.} \emph{et~al.}
\newblock \bibinfo{title}{Extreme ultraviolet high-harmonic spectroscopy of
  solids}.
\newblock \emph{\bibinfo{journal}{Nature}} \textbf{\bibinfo{volume}{521}},
  \bibinfo{pages}{498} (\bibinfo{year}{2015}).

\bibitem{garg2016multi}
\bibinfo{author}{Garg, M.} \emph{et~al.}
\newblock \bibinfo{title}{Multi-petahertz electronic metrology}.
\newblock \emph{\bibinfo{journal}{Nature}} \textbf{\bibinfo{volume}{538}},
  \bibinfo{pages}{359} (\bibinfo{year}{2016}).

\bibitem{banks2017dynamical}
\bibinfo{author}{Banks, H.~B.} \emph{et~al.}
\newblock \bibinfo{title}{Dynamical birefringence: electron-hole recollisions
  as probes of berry curvature}.
\newblock \emph{\bibinfo{journal}{Physical Review X}}
  \textbf{\bibinfo{volume}{7}}, \bibinfo{pages}{041042} (\bibinfo{year}{2017}).

\bibitem{luu2018measurement}
\bibinfo{author}{Luu, T.~T.} \& \bibinfo{author}{W{\"o}rner, H.~J.}
\newblock \bibinfo{title}{Measurement of the berry curvature of solids using
  high-harmonic spectroscopy}.
\newblock \emph{\bibinfo{journal}{Nature Communications}}
  \textbf{\bibinfo{volume}{9}}, \bibinfo{pages}{1--6} (\bibinfo{year}{2018}).

\bibitem{bauer2018high}
\bibinfo{author}{Bauer, D.} \& \bibinfo{author}{Hansen, K.~K.}
\newblock \bibinfo{title}{High-harmonic generation in solids with and without
  topological edge states}.
\newblock \emph{\bibinfo{journal}{Physical Review Letters}}
  \textbf{\bibinfo{volume}{120}}, \bibinfo{pages}{177401}
  (\bibinfo{year}{2018}).

\bibitem{jurss2019high}
\bibinfo{author}{J{\"u}r{\ss}, C.} \& \bibinfo{author}{Bauer, D.}
\newblock \bibinfo{title}{High-harmonic generation in su-schrieffer-heeger
  chains}.
\newblock \emph{\bibinfo{journal}{Physical Review B}}
  \textbf{\bibinfo{volume}{99}}, \bibinfo{pages}{195428}
  (\bibinfo{year}{2019}).

\bibitem{murakami2018high}
\bibinfo{author}{Murakami, Y.}, \bibinfo{author}{Eckstein, M.} \&
  \bibinfo{author}{Werner, P.}
\newblock \bibinfo{title}{High-harmonic generation in mott insulators}.
\newblock \emph{\bibinfo{journal}{Physical Review Letters}}
  \textbf{\bibinfo{volume}{121}}, \bibinfo{pages}{057405}
  (\bibinfo{year}{2018}).

\bibitem{reimann2018subcycle}
\bibinfo{author}{Reimann, J.} \emph{et~al.}
\newblock \bibinfo{title}{Subcycle observation of lightwave-driven dirac
  currents in a topological surface band}.
\newblock \emph{\bibinfo{journal}{Nature}} \textbf{\bibinfo{volume}{562}},
  \bibinfo{pages}{396--400} (\bibinfo{year}{2018}).

\bibitem{imai2020high}
\bibinfo{author}{Imai, S.}, \bibinfo{author}{Ono, A.} \&
  \bibinfo{author}{Ishihara, S.}
\newblock \bibinfo{title}{High harmonic generation in a correlated electron
  system}.
\newblock \emph{\bibinfo{journal}{Physical Review Letters}}
  \textbf{\bibinfo{volume}{124}}, \bibinfo{pages}{157404}
  (\bibinfo{year}{2020}).

\bibitem{baykusheva2020strong}
\bibinfo{author}{Baykusheva, D.} \emph{et~al.}
\newblock \bibinfo{title}{Strong-field physics in three-dimensional topological
  insulators}.
\newblock \emph{\bibinfo{journal}{Physical Review A}}
  \textbf{\bibinfo{volume}{103}}, \bibinfo{pages}{023101}
  (\bibinfo{year}{2021}).

\bibitem{bai2020high}
\bibinfo{author}{Bai, Y.} \emph{et~al.}
\newblock \bibinfo{title}{High-harmonic generation from topological surface
  states}.
\newblock \emph{\bibinfo{journal}{Nature Physics}}
  \textbf{\bibinfo{volume}{17}}, \bibinfo{pages}{311--315}
  (\bibinfo{year}{2021}).

\bibitem{borsch2020super}
\bibinfo{author}{Borsch, M.} \emph{et~al.}
\newblock \bibinfo{title}{Super-resolution lightwave tomography of electronic
  bands in quantum materials}.
\newblock \emph{\bibinfo{journal}{Science}} \textbf{\bibinfo{volume}{370}},
  \bibinfo{pages}{1204--1207} (\bibinfo{year}{2020}).

\bibitem{mrudul2020high_new}
\bibinfo{author}{Mrudul, M.~S.} \& \bibinfo{author}{Dixit, G.}
\newblock \bibinfo{title}{High-harmonic generation from monolayer and bilayer
  graphene}.
\newblock \emph{\bibinfo{journal}{Physical Review B}}
  \textbf{\bibinfo{volume}{103}}, \bibinfo{pages}{094308}
  (\bibinfo{year}{2021}).

\bibitem{beenakker2013search}
\bibinfo{author}{Beenakker, C. W.~J.}
\newblock \bibinfo{title}{Search for majorana fermions in superconductors}.
\newblock \emph{\bibinfo{journal}{Annu. Rev. Condens. Matter Phys.}}
  \textbf{\bibinfo{volume}{4}}, \bibinfo{pages}{113--136}
  (\bibinfo{year}{2013}).

\bibitem{sato2017topological}
\bibinfo{author}{Sato, M.} \& \bibinfo{author}{Ando, Y.}
\newblock \bibinfo{title}{Topological superconductors: a review}.
\newblock \emph{\bibinfo{journal}{Reports on Progress in Physics}}
  \textbf{\bibinfo{volume}{80}}, \bibinfo{pages}{076501}
  (\bibinfo{year}{2017}).

\bibitem{majorana}
\bibinfo{author}{Majorana, E.}
\newblock \bibinfo{title}{A symmetric theory of electrons and positrons}.
\newblock \emph{\bibinfo{journal}{Nuovo Cim.}} \textbf{\bibinfo{volume}{14}},
  \bibinfo{pages}{171--184} (\bibinfo{year}{1937}).

\bibitem{elliott2015colloquium}
\bibinfo{author}{Elliott, S.~R.} \& \bibinfo{author}{Franz, M.}
\newblock \bibinfo{title}{Colloquium: Majorana fermions in nuclear, particle,
  and solid-state physics}.
\newblock \emph{\bibinfo{journal}{Reviews of Modern Physics}}
  \textbf{\bibinfo{volume}{87}}, \bibinfo{pages}{137} (\bibinfo{year}{2015}).

\bibitem{kitaev2003fault}
\bibinfo{author}{Kitaev, A.~Y.}
\newblock \bibinfo{title}{Fault-tolerant quantum computation by anyons}.
\newblock \emph{\bibinfo{journal}{Annals of Physics}}
  \textbf{\bibinfo{volume}{303}}, \bibinfo{pages}{2--30}
  (\bibinfo{year}{2003}).

\bibitem{alicea2012new}
\bibinfo{author}{Alicea, J.}
\newblock \bibinfo{title}{New directions in the pursuit of majorana fermions in
  solid state systems}.
\newblock \emph{\bibinfo{journal}{Reports on Progress in Physics}}
  \textbf{\bibinfo{volume}{75}}, \bibinfo{pages}{076501}
  (\bibinfo{year}{2012}).

\bibitem{nayak2008non}
\bibinfo{author}{Nayak, C.}, \bibinfo{author}{Simon, S.~H.},
  \bibinfo{author}{Stern, A.}, \bibinfo{author}{Freedman, M.} \&
  \bibinfo{author}{Sarma, S.~D.}
\newblock \bibinfo{title}{Non-abelian anyons and topological quantum
  computation}.
\newblock \emph{\bibinfo{journal}{Reviews of Modern Physics}}
  \textbf{\bibinfo{volume}{80}}, \bibinfo{pages}{1083} (\bibinfo{year}{2008}).

\bibitem{sarma2015majorana}
\bibinfo{author}{Sarma, S.~D.}, \bibinfo{author}{Freedman, M.} \&
  \bibinfo{author}{Nayak, C.}
\newblock \bibinfo{title}{Majorana zero modes and topological quantum
  computation}.
\newblock \emph{\bibinfo{journal}{npj Quantum Information}}
  \textbf{\bibinfo{volume}{1}}, \bibinfo{pages}{1--13} (\bibinfo{year}{2015}).

\bibitem{mourik2012signatures}
\bibinfo{author}{Mourik, V.} \emph{et~al.}
\newblock \bibinfo{title}{Signatures of majorana fermions in hybrid
  superconductor-semiconductor nanowire devices}.
\newblock \emph{\bibinfo{journal}{Science}} \textbf{\bibinfo{volume}{336}},
  \bibinfo{pages}{1003--1007} (\bibinfo{year}{2012}).

\bibitem{das2012zero}
\bibinfo{author}{Das, A.} \emph{et~al.}
\newblock \bibinfo{title}{Zero-bias peaks and splitting in an al--inas nanowire
  topological superconductor as a signature of majorana fermions}.
\newblock \emph{\bibinfo{journal}{Nature Physics}}
  \textbf{\bibinfo{volume}{8}}, \bibinfo{pages}{887--895}
  (\bibinfo{year}{2012}).

\bibitem{albrecht2016exponential}
\bibinfo{author}{Albrecht, S.~M.} \emph{et~al.}
\newblock \bibinfo{title}{Exponential protection of zero modes in majorana
  islands}.
\newblock \emph{\bibinfo{journal}{Nature}} \textbf{\bibinfo{volume}{531}},
  \bibinfo{pages}{206--209} (\bibinfo{year}{2016}).

\bibitem{kim2018toward}
\bibinfo{author}{Kim, H.} \emph{et~al.}
\newblock \bibinfo{title}{Toward tailoring majorana bound states in
  artificially constructed magnetic atom chains on elemental superconductors}.
\newblock \emph{\bibinfo{journal}{Science advances}}
  \textbf{\bibinfo{volume}{4}}, \bibinfo{pages}{eaar5251}
  (\bibinfo{year}{2018}).

\bibitem{wang2018evidence}
\bibinfo{author}{Wang, D.} \emph{et~al.}
\newblock \bibinfo{title}{Evidence for majorana bound states in an iron-based
  superconductor}.
\newblock \emph{\bibinfo{journal}{Science}} \textbf{\bibinfo{volume}{362}},
  \bibinfo{pages}{333--335} (\bibinfo{year}{2018}).

\bibitem{jack2019observation}
\bibinfo{author}{J{\"a}ck, B.} \emph{et~al.}
\newblock \bibinfo{title}{Observation of a majorana zero mode in a
  topologically protected edge channel}.
\newblock \emph{\bibinfo{journal}{Science}} \textbf{\bibinfo{volume}{364}},
  \bibinfo{pages}{1255--1259} (\bibinfo{year}{2019}).

\bibitem{lutchyn2018majorana}
\bibinfo{author}{Lutchyn, R.~M.} \emph{et~al.}
\newblock \bibinfo{title}{Majorana zero modes in superconductor--semiconductor
  heterostructures}.
\newblock \emph{\bibinfo{journal}{Nature Reviews Materials}}
  \textbf{\bibinfo{volume}{3}}, \bibinfo{pages}{52--68} (\bibinfo{year}{2018}).

\bibitem{wang2020evidence}
\bibinfo{author}{Wang, Z.} \emph{et~al.}
\newblock \bibinfo{title}{Evidence for dispersing 1d majorana channels in an
  iron-based superconductor}.
\newblock \emph{\bibinfo{journal}{Science}} \textbf{\bibinfo{volume}{367}},
  \bibinfo{pages}{104--108} (\bibinfo{year}{2020}).

\bibitem{kitaev2001unpaired}
\bibinfo{author}{Kitaev, A.}
\newblock \bibinfo{title}{Unpaired majorana fermions in quantum wires}.
\newblock \emph{\bibinfo{journal}{Physics-Uspekhi}}
  \textbf{\bibinfo{volume}{44}}, \bibinfo{pages}{131} (\bibinfo{year}{2001}).

\bibitem{Ohm_Hassler_2014}
\bibinfo{author}{{Ohm}, C.} \& \bibinfo{author}{{Hassler}, F.}
\newblock \bibinfo{title}{{Majorana fermions coupled to electromagnetic
  radiation}}.
\newblock \emph{\bibinfo{journal}{New Journal of Physics}}
  \textbf{\bibinfo{volume}{16}}, \bibinfo{pages}{015009}
  (\bibinfo{year}{2014}).

\bibitem{DeGennes_book}
\bibinfo{author}{De~Gennes, P.~G.}
\newblock \emph{\bibinfo{title}{Superconductivity of Metals and Alloys}}.
\newblock Advanced book classics (\bibinfo{publisher}{Perseus},
  \bibinfo{address}{Cambridge, MA}, \bibinfo{year}{1999}).
\newblock \urlprefix\url{https://cds.cern.ch/record/566105}.

\bibitem{zachmann2013ultrafast}
\bibinfo{author}{Zachmann, M.} \emph{et~al.}
\newblock \bibinfo{title}{Ultrafast terahertz-field-induced dynamics of
  superconducting bulk and quasi-1d samples}.
\newblock \emph{\bibinfo{journal}{New Journal of Physics}}
  \textbf{\bibinfo{volume}{15}}, \bibinfo{pages}{055016}
  (\bibinfo{year}{2013}).

\bibitem{silva2018high}
\bibinfo{author}{Silva, R. E.~F.}, \bibinfo{author}{Blinov, I.~V.},
  \bibinfo{author}{Rubtsov, A.~N.}, \bibinfo{author}{Smirnova, O.} \&
  \bibinfo{author}{Ivanov, M.}
\newblock \bibinfo{title}{High-harmonic spectroscopy of ultrafast many-body
  dynamics in strongly correlated systems}.
\newblock \emph{\bibinfo{journal}{Nature Photonics}}
  \textbf{\bibinfo{volume}{12}}, \bibinfo{pages}{266} (\bibinfo{year}{2018}).

\end{thebibliography}

\end{document}